\documentclass{emulateapj}

\slugcomment{}

\shorttitle{Alfv\'{e}n Waves in the Lower Solar Atmosphere}
\shortauthors{D.B. Jess et al.}

\begin{document}

\title{Alfv\'{e}n Waves in the Lower Solar Atmosphere}

\author{D. B. Jess}
\affil{Astrophysics Research Centre, School of Mathematics and Physics, Queen's University, Belfast, BT7~1NN,
Northern Ireland, U.K.}
\affil{}
\affil{NASA Goddard Space Flight Center, Solar Physics Laboratory, Code 671, Greenbelt, MD 20771, USA}
\email{d.jess@qub.ac.uk}

\author{M. Mathioudakis}
\affil{Astrophysics Research Centre, School of Mathematics and Physics, Queen's University, Belfast, BT7~1NN,
Northern Ireland, U.K.}

\author{R. Erd\'{e}lyi}
\affil{SP$^{2}$RC, Department of Applied Mathematics, The University of Sheffield, Sheffield, S3 7RH,
England, U.K.}

\author{P. J. Crockett and F. P. Keenan}
\affil{Astrophysics Research Centre, School of Mathematics and Physics, Queen's University, Belfast, BT7~1NN,
Northern Ireland, U.K.}

\and

\author{D. J. Christian}
\affil{Department of Physics and Astronomy, California State University Northridge, 18111 Nordhoff Street, Northridge, 
CA 91330, USA.}

\author{~}
\affil{~}



\begin{abstract}
The flow of energy through the solar atmosphere and the heating of the Sun's outer regions are
still not understood. Here, we report the detection of oscillatory phenomena associated with a large
bright-point group that is 430,000 square kilometers in area and located near the solar disk
center. Wavelet analysis reveals full-width half-maximum oscillations with periodicities ranging
from 126 to 700 seconds originating above the bright point and significance levels exceeding
99\%. These oscillations, 2.6 kilometers per second in amplitude, are coupled with chromospheric
line-of-sight Doppler velocities with an average blue shift of 23 kilometers per second. A lack
of cospatial intensity oscillations and transversal displacements rules out the presence of
magneto-acoustic wave modes. The oscillations are a signature of Alfv\'{e}n waves produced by a
torsional twist of $\pm$22 degrees. A phase shift of 180 degrees across the diameter of the bright point
suggests that these torsional Alfv\'{e}n oscillations are induced globally throughout the entire
brightening. The energy flux associated with this wave mode is sufficient to heat the solar corona.

\end{abstract}

\keywords{Published in Science 20-March-2009}

\section{Report}
\label{report}

Solar observations from both ground-based
and spaceborne facilities show that a wide
range of magneto-acoustic waves (1, 2) propagate
throughout the solar atmosphere. However,
the energy they carry to the outer solar atmosphere
is not sufficient to heat it (3). Alfv\'{e}n waves (pure
magnetic waves), which are incompressible and
can penetrate through the stratified solar atmosphere
without being reflected (4), are the most promising
wave mechanism to explain the heating of the
Sun's outer regions.

\begin{figure*}
\epsscale{1.0}
\plotone{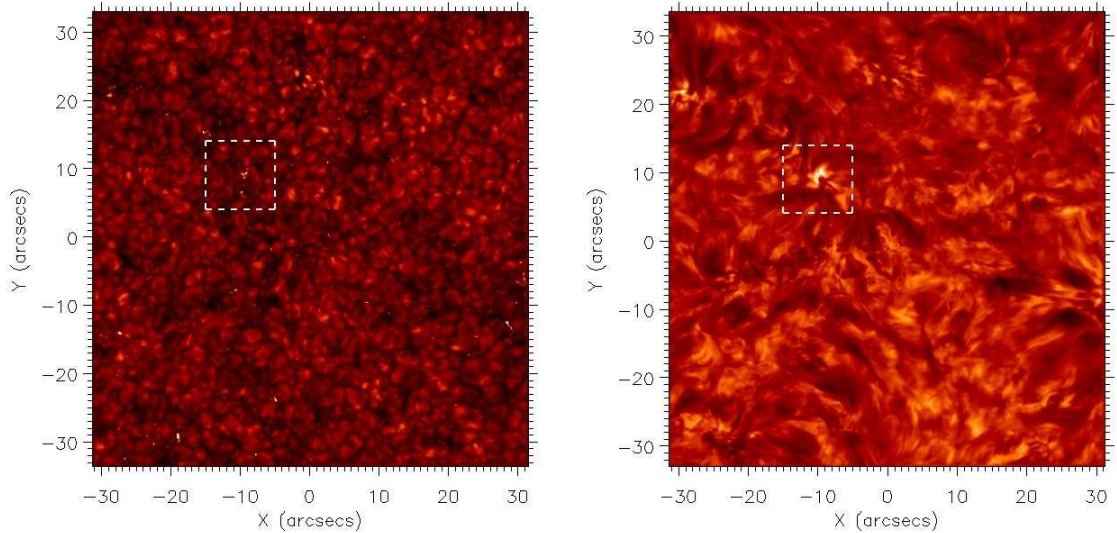}
\caption{Simultaneous images in the (left) H$\alpha$
continuum (photosphere)
and (right) H$\alpha$ core (chromosphere)
obtained with
the SST. The conglomeration
of bright points
within the region we investigated
is denoted by
a square of dashed lines.
The scale is in heliocentric
coordinates where
1 arc sec $\approx$ 725~km.}
\label{f1}
\end{figure*}

However, it has been suggested that their previous
detection in the solar corona (5) and upper
chromosphere (6) is inconsistent with magnetohydrodynamic
(MHD) wave theory (7, 8). These
observations are best interpreted as a guided-kink
magneto-acoustic mode, whereby the observational
signatures are usually swaying, transversal,
periodic motions of the magnetic flux tubes
(7, 9). Numerical simulations (10) show that subsurface
acoustic drivers and fast magneto-sonic
kink waves (11, 12) can convert energy into upwardly
propagating Alfv\'{e}n waves, which are
emitted from the solar surface. These numerical
simulations are also in agreement with current
analytical studies. In particular, it has been shown
that footpoint motions in an axially symmetric
system can excite torsional Alfv\'{e}n waves (13).
Other Alfv\'{e}n wave modes may exist, although
these are normally coupled to magneto-sonic MHD
waves (14). In the solar atmosphere, magnetic
field lines clump into tight bundles, forming flux
tubes. Alfv\'{e}n waves in flux tubes could manifest
as torsional oscillations (7) that create simultaneous
blue and red shifts, leading to the non-thermal
broadening of any isolated line profile,
and should thus be observed as full-width half-maximum
(FWHM) oscillations (15). A promising
location for the detection of Alfv\'{e}n waves is 
in the lower solar atmosphere, where they can be
generated by the overshooting of convective motions
in the photosphere (16). Here, we report the
detection of substantially blue-shifted plasma and
FWHM oscillations originating in a large conglomeration
of magnetic bright points.

We used the Swedish Solar Telescope (SST) to
image a 68-by-68-arc sec region on the solar surface
positioned near the disk center. Using the Solar
Optical Universal Polarimeter (SOUP) (17) and
high-order adaptive optics (18), we obtained narrowband
images across the H$\alpha$ absorption profile centered
at 6562.8~\AA. We observed with a cadence of
0.03~s to obtain 89~min of uninterrupted data. Because
SOUP is tunable, we sampled the complete
H$\alpha$ line profile using seven discrete steps. The wavelength
intervals we chose became increasingly narrow
toward the line core in order to enable an
accurate determination of the line characteristics,
such as Doppler velocities, FWHMs, and intensities.
Our images have a sampling of 0.068 arc sec
per pixel, which corresponds to $\approx$110-km resolution
(two pixels) on the solar surface.

By using the multi-object multi-frame blind deconvolution
(MOMFBD) (19) image restoration
technique to remove the small-scale atmospheric
distortions present in the data, we achieved an effective
cadence of 63~s for a full line profile. We
acquired 85 complete scans across the H$\alpha$ line profile
in addition to 595 simultaneous continuum images.
For each of the 85 profile scans, every pixel
of the 1024-by-1024-pixel$^{2}$ charge-coupled device
contains information acquired at a particular wavelength
position. Therefore, we obtained a total of
$8.9 \times 10^{7}$ individual H$\alpha$ absorption profiles, covering
the full 68-by-68-arc sec field of view, during
the 89-min duration of the data set. We fit a
Gaussian distribution to each of the observed H$\alpha$
profiles to obtain values for the integrated intensity
and FWHM. To determine the line-of-sight
velocity, we compared each measured central wavelength
position with the rest-frame H$\alpha$ profile core
at 6562.8~\AA. We created time series for intensity,
line-of-sight velocity, FWHM, and wavelength-integrated
data cubes and used fast Fourier transform
and wavelet routines to analyze them.

The SST field of view shows a range of features,
including pores, exploding granules, and a multitude
of bright points (Fig. 1), a large conglomeration of
which is located at heliocentric coordinates ($-10$ arc
sec, $10$ arc sec) or N07E01 in the solar north-south-east-west 
coordinate system. We selected a 10-by-10-arc 
sec box surrounding the bright-point group
(BPG), which occupies an area of 430,000 km$^{2}$, for
further investigation. The line-of-sight Doppler velocities
associated with this BPG show blue shifts
with an average value of 23 km~s$^{-1}$. There is no
evidence of periodic trends in either intensity or line-of-sight velocity; 
the intensity of the BPG is constant,
with minimal variation during its 53-min lifetime.

Wavelet analysis shows that FWHM oscillations
with significance levels exceeding 99\% occur
within the spatially averaged BPG (Fig. 2). We
detected FWHM oscillations as low as the Nyquist
period (126~s) throughout the duration of the data
set, with the strongest detected power originating in
the 400-to-500-s interval. These oscillations are located
directly above the large BPG, encompassing a
near-circular shape that is cospatial with the detected
Doppler velocities, and are apparent in all
FWHM time series. This shows that powerful coherent
periodicities are present throughout the
surface of the BPG. We detected oscillations until
the BPG fragmented into a series of smaller bright
points after 3150~s.

\begin{figure}
\epsscale{1.0}
\plotone{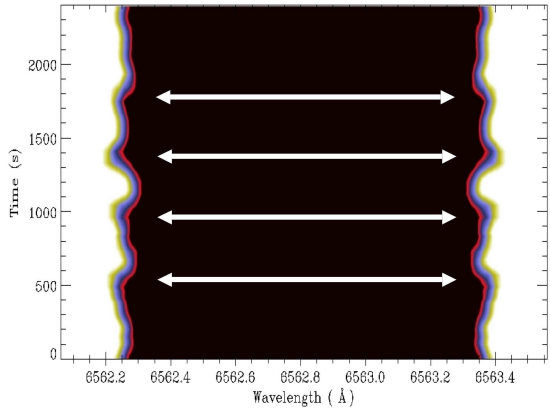}
\caption{A wavelength-versus-time plot of the H$\alpha$
profile showing the variation of line width at
FWHM as a function of time. The arrows indicate
the positions of maximum amplitude of a 420-s
periodicity associated with the bright-point
group located at ($-10$ arc~sec, $10$ arc~sec) in
Fig. 1. The torsional motion of the Alfv\'{e}nic
perturbations creates nonthermal broadening
that is visible in the H$\alpha$ line profile. The peak-to-peak velocity 
is $\approx$3.0~km~s$^{-1}$ ($\approx$65~m\AA). For
an inclination angle of 35$^{\circ}$, the absolute velocity
amplitude is $\approx$2.6~km~s$^{-1}$.}
\label{f2}
\end{figure}

Numerical simulations based on three-dimensional
magnetoconvection show that the
bright points that were observed in the wing of
the H$\alpha$ line profile correspond to magnetic field
concentrations measured in kilogauss in the photosphere
(20). The canopy structure seen in the H$\alpha$
core images reveals a wealth of flux-tube structures,
with many securing anchor positions in the photosphere
directly above the BPG (Fig. 1). The coincidence
of bright-point structures with high magnetic
field concentrations implies that MHD waves are
likely to be present (21). However, the chromospheric
brightening is of much larger physical size
than the underlying photospheric BPG. Because the
observations were made very close to the center of
the solar disk, an increase in physical size between
the photosphere and the chromosphere can be interpreted
as an expansion of the photospheric flux-tube
bundle as a function of atmospheric height
(22). A comparison of the maximum diameter of
the bright point at each height in the atmosphere
suggests an expansion of $\approx$1300~km; a height separation
of $\approx$1000~km and a symmetric expansion
around the bright-point center suggest a flux-tube
expansion angle of $\approx$33$^{\circ}$. Additionally, an offset of
$\approx$700~km between the centres of the BPG at photospheric
and chromospheric heights suggests a magnetic
flux-tube tilt angle of $\approx$35$^{\circ}$ from the vertical.

Alfv\'{e}nic fluxes are predicted to be at their
strongest in the regime of high magnetic field
strength and moderately inclined waveguides (10).
Because of their incompressibility, they exhibit no
periodic intensity perturbations. Thus, the observational
signature of a torsional Alfv\'{e}n wave propagating
with a velocity component along the observer's
line of sight will arise from its torsional velocities
on small spatial scales (8). These torsional velocities
are responsible for the FWHM oscillations we
observed (Fig. 3). The line-of-sight velocity amplitude
of $\approx$1.5~km~s$^{-1}$ and the inclination angle of
$\approx$35$^{\circ}$ indicate an absolute Alfv\'{e}nic perturbation
amplitude of $\approx$2.6~km~s$^{-1}$. Because the circumference
of the photospheric bright point [where
torsional Alfv\'{e}n waves are believed to be generated
(16)] is on the order of 2800~km (55 pixels), a
torsional twist of $\pm$22$^{\circ}$ is sufficient to generate the
observed wave motion.

The moderate inclination angle of the flux tubes,
coupled with the detection of substantially blue-shifted
material and strong FWHM oscillations, is
evidence of the presence of torsional Alfv\'{e}n waves.
For a typical photospheric internal waveguide electron
density (23) of $n_{e}^{i} \approx 10^{16}$~cm$^{-3}$ and a magnetic
field strength (20) of 1000~G, the Alfv\'{e}n speed
within a cylindrical flux tube is estimated (14) to be
$\approx$22~km~s$^{-1}$. This value is above the speed of sound
in the upper photosphere/lower chromosphere
(24) and is consistent with the blue-shift velocity
we determined.

\begin{figure}
\epsscale{1.0}
\plotone{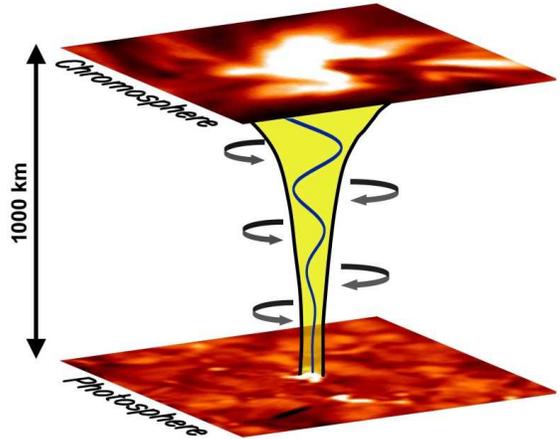}
\caption{Expanding magnetic
flux tube sandwiched
between photospheric and
chromospheric intensity
images obtained with the
SST, undergoing a torsional
Alfv\'{e}nic perturbation
and generating a wave
that propagates longitudinally
in the vertical direction.
At a given position
along the flux tube, the
Alfv\'{e}nic displacements are
torsional oscillations that
remain perpendicular to
the direction of propagation
and magnetic field
outlining constant magnetic
surfaces. The largest
FWHM will be produced
when the torsional velocity
is at its maximum (at
zero displacement from the equilibrium position). The figure is not to scale.}
\label{f3}
\end{figure}

We took a slice through the center of the bright
point and analyzed the stability of opposite edges
of the BPG as a function of time. This was performed
by examining any displacements of the
BPG from its initial position at the start of the observing
sequence (fig. S1). The bright point moves
less than one pixel during the first 3150~s. As the
BPG begins to fragment after 3150~s, the motions
of the bright-point edges increase substantially.
However, we did not find periodic motions of the
bright point, particularly during the initial 3150~s
when the FWHM oscillations were detected. A
magneto-acoustic wave mode would produce observable 
periodicities in intensity, similar to those
caused by the periodic contractions when viewed
along the flux tube, that are associated with sausage-mode
waves (25). A sausage-mode wave is caused
by the axially symmetric expansion and contraction
of magnetic flux tubes (14). Kink-mode oscillations
are generated through a bulk motion, whereby the
whole flux tube is displaced from its original position
in a periodic fashion. Therefore, magneto-acoustic
waves cannot explain our observations.

A torsional Alfv\'{e}nic perturbation should produce
a FWHM oscillation that is 180$^{\circ}$ out of phase
at opposite boundaries of the waveguide (8).
The relative time-averaged oscillatory phase as a
function of distance across the $\approx$2200-km diameter
of the bright point shows that opposite sides of
the bright point display oscillatory phenomena
that are indeed 180$^{\circ}$ out of phase (Fig. 4). This
is consistent with current torsional Alfv\'{e}nic wave
models (26).

\begin{figure}
\epsscale{1.0}
\plotone{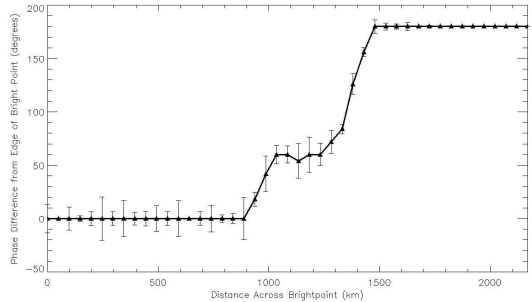}
\caption{Average phase difference of FWHM oscillations plotted as a function of distance across the diameter
of the bright point. The triangles denote the locations where a measurement is made, and the error bars
indicate the phase variance in the temporal domain. The phase at 0~km is used as a reference, with all other
phases plotted relative to this value. Spatial coherence of FWHM oscillations across the BPG ranges between
90 and 100\%, suggesting that the BPG is acting as a coherent waveguide. The phase difference increases
around the midpoint of the BPG with opposite sides of the waveguide, indicating out-of-phase oscillatory
phenomena, which is as predicted for a torsional Alfv\'{e}nic perturbation.}
\label{f4}
\end{figure}

We estimated the energy flux of the observed
waves using $E = \rho v^{2} v_{A}$, where $\rho$ is the mass density
of the flux-tube, $v$ is the observed velocity amplitude,
and $v_{A}$ is the Alfv\'{e}n speed (6). For a mass
density of $\rho \approx 1 \times 10^{-6}$~kg~m$^{-3}$, derived from a
quiet-Sun chromospheric model (23), the energy
flux in the chromosphere is $E \approx 15000$~W~m$^{-2}$. At
any one time, it is estimated (27) that at least 1.6\%
of the solar surface is covered by BPGs similar to
that presented here. Thus, combining the energy
carried by similar BPGs over the entire solar surface
produces a global average of 240~W~m$^{-2}$.
Alfv\'{e}n waves with an energy flux of $\approx$100~W~m$^{-2}$
are believed to be vigorous enough to heat the localized
corona or to launch the solar wind when
their energy is thermalized (6, 28). Therefore, a transmission
coefficient of $\approx$42\% through the thin transition
region will provide sufficient energy to heat
the entire corona. Regions containing highly magnetic
structures, such as bright points, should possess
even higher mass densities (29). In this regime,
the energy flux available to heat the corona will
be substantially higher than the minimum value
required to sustain localized heating.

\section{References and Notes}
\begin{enumerate}
\item{Magneto-acoustic waves, normally classified as fast and
slow, are waves of acoustic origin whose properties are
modified by the presence of a magnetic field.}
\item{V. M. Nakariakov, E. Verwichte, Living Rev. Sol. Phys. 2, 3 (2005).}
\item{A. Fossum, M. Carlsson, Nature 435, 919 (2005).}
\item{L. Ofman, Astrophys. J. 568, L135 (2002).}
\item{S. Tomczyk et al., Science 317, 1192 (2007).}
\item{B. De Pontieu et al., Science 318, 1574 (2007).}
\item{R. Erd\'{e}lyi, V. Fedun, Science 318, 1572 (2007).}
\item{T. Van Doorsselaere, V. Nakariakov, E. Verwichte,
Astrophys. J. 676, L73 (2008).}
\item{V. Kukhianidze, T. V. Zaqarashvili, E. Khutsishvili,
Astron. Astrophys. 449, L35 (2006).}
\item{P. S. Cally, M. Goossens, Sol. Phys. 251, 251 (2008).}
\item{M. Goossens, I. Arregui, J. L. Ballester, T. J. Wang,
Astron. Astrophys. 484, 851 (2008).}
\item{W. J. Tirry, M. Goossens, Astrophys. J. 471, 501 (1996).}
\item{M. S. Ruderman, D. Berghmans, M. Goossens, S. Poedts,
Astron. Astrophys. 320, 305 (1997).}
\item{P. M. Edwin, B. Roberts, Sol. Phys. 88, 179 (1983).}
\item{T. V. Zaqarashvili, Astron. Astrophys. 399, L15 (2003).}
\item{J. Vranjes, S. Poedts, B. P. Pandey, B. De Pontieu,
Astron. Astrophys. 478, 553 (2008).}
\item{A. M. Title, W. J. Rosenberg, Opt. Eng. 20, 815 (1981).}
\item{G. B. Scharmer, P. M. Dettori, M. G. Lofdahl, M. Shand,
Proc. SPIE 4853, 370 (2003).}
\item{M. van Noort, L. H. M. Rouppe van der Voort, M. G. Lofdahl,
Sol. Phys. 228, 191 (2005).}
\item{J. Leenaarts, R. J. Rutten, P. Sütterlin, M. Carlsson,
H. Uitenbroek, Astron. Astrophys. 449, 1209 (2006).}
\item{W. Kalkofen, Astrophys. J. 486, L145 (1997).}
\item{S. K. Solanki, W. Finsterle, I. Ruedi, W. Livingston,
Astron. Astrophys. 347, L27 (1999).}
\item{J. E. Vernazza, E. H. Avrett, R. Loeser, Astrophys. J. Suppl.
45, 635 (1981).}
\item{B. S\'{a}nchez-Andrade N\~{u}no, N. Bello Gonz\'{a}lez, J. Blanco
Rodriguez, F. Kneer, K. G. Puschmann, Astron. Astrophys.
486, 577 (2008).}
\item{V. M. Nakariakov, V. F. Melnikov, V. E. Reznikova,
Astron. Astrophys. 412, L7 (2003).}
\item{P. Copil, Y. Voitenko, M. Goossens, Astron. Astrophys.
478, 921 (2008).}
\item{D. S. Brown, C. E. Parnell, E. E. Deluca, L. Golub,
R. A. McMullen, Sol. Phys. 201, 305 (2001).}
\item{A. Verdini, M. Velli, Astrophys. J. 662, 669 (2007).}
\item{D. P\'{er}ez-Su\'{a}rez, R. C. Maclean, J. G. Doyle, M. S. Madjarska,
Astron. Astrophys. 492, 575 (2008).}
\item{D.B.J. is supported by a Northern Ireland Department
for Employment and Learning studentship and
thanks NASA Goddard Space Flight Center for a
Co-operative Award in Science and Technology
studentship. R.E. thanks M. K\'{e}ray for encouragement
and is grateful to NSF, Hungary (Orsz\'{a}gos Tudom\'{a}nyos
Kutat\'{a}si Alapprogram, ref. no. K67746), for
financial support. F.P.K. is grateful to the Atomic
Weapons Establishment-Aldermaston for the
award of a William Penney Fellowship. The SST is
operated on the island of La Palma by the Institute
for Solar Physics of the Royal Swedish Academy of
Sciences in the Spanish Observatorio del Roque de los
Muchachos of the Instituto de Astrof\'{i}sica de Canarias.
These observations have been funded by the Optical
Infrared Coordination network, an international
collaboration supported by the Research Infrastructures
Programme of the European Commission's Sixth
Framework Programme. This work is supported by the
Science and Technology Facilities Council, and we thank
L. H. M. Rouppe van der Voort for help with MOMFBD
image processing.}
\end{enumerate}

\clearpage




\clearpage


\clearpage


\clearpage


\clearpage

\end{document}